\documentstyle[12pt,amsfonts,amssymbols]{article}

\makeatletter

\makeatletter

\@addtoreset{equation}{section}
\def\section{\@startsection {section}{1}{\z@}{-3.5ex plus -1ex minus
 -.2ex}{2.3ex plus .2ex}{\large\bf}}
\def\subsection{\@startsection{subsection}{2}{\z@}{-3.25ex plus
 -1ex minus -.2ex}{1.5ex plus .2ex}{\sc}}
\fnsymbol{footnote}

\advance \voffset by -1.0in
\advance \hoffset by -0.6in
\def\bZ{\mathbb{Z}}
\def\bR{\mathbb{R}}
\def\bea{\begin{eqnarray}}
\def\eea{\end{eqnarray}}
\begin{document}
\baselineskip 18pt
\parskip 3pt
\begin{flushright}

ITFA-97-38

hep-th/9710182

\end{flushright}

\vspace {2cm}
 
\begin{center}

{\LARGE Non-abelian Dyons\footnote{talk given at ``Solitons '97'', 
Kingston, Ontario , July 1997 }}

\vspace{1cm}
\baselineskip 24pt

{\large 
B.J. Schroers\footnote{e-mail: {\tt schroers@phys.uva.nl}}
\\
Instituut voor Theoretische Fysica
\\
Valckenierstraat 65
\\
1018 XE Amsterdam
\\
The Netherlands
}

\vspace{1.5cm}

{\bf Abstract}

\end{center}
\baselineskip 15pt

{\small 
\noindent 
The dyonic quantum states of 
magnetic monopoles in Yang-Mills-Higgs theory with a non-abelian 
unbroken gauge group   display a subtle interplay between 
magnetic and electric properties. This is described in detail
in  the  theory with the gauge group  $SU(3)$ broken to $U(2)$
and shown to be captured by the representation theory of the 
semi-direct product $U(2) \ltimes {\bR}^4$.
The implications of this observation for the fusion rules and 
electric-magnetic duality properties of dyonic states are pointed out. }

\vspace{2cm}

\vfill 

\pagebreak

\baselineskip 18pt

\section{Outline of the problem } 
One of the most exciting applications of the soliton concept is the
physics of elementary particles. This application is  very natural
in view of the particle-like properties of  solitons, but it requires that 
one complements the classical theory of solitons by quantum concepts.
A fully (3+1)-dimensional model in which both the classical and 
the (semi-) quantised properties of solitons are particularly well-studied
is Yang-Mills-Higgs (YMH) theory  in the Prasad-Sommerfield limit.
The soliton solutions in this theory carry magnetic charge and
are called magnetic monopoles. Naturally most is known about 
the monopole solutions in the theory with the simplest gauge
group, namely $SU(2)$ broken to $U(1)$, see \cite{AH}
for a review. They provide  a paradigmatic
example for classical and quantised soliton properties and
also inspired  Montonen and  Olive to formulate
  the first precise duality conjecture in a
(3+1)-dimensional field theory \cite{MO}.

In this talk I want to discuss  the qualitatively new questions
that arise when one tries to understand the classical and quantum properties
of monopoles in  YMH theory with non-abelian
 unbroken gauge group.
The talk is based on research carried out jointly with
Sander Bais and reported in  the paper \cite{BS}.
To present the questions I want to address in as sharp a light as possible,
let me briefly review salient features of  monopoles 
in $SU(2)$ YMH theory broken to $U(1)$.

\begin{enumerate}
\item 
The solitons carry an integer topological charge which  one may
interpret as the particle number. 
This charge is an element of the homotopy
group $\Pi_2(SU(2)/U(1))=\bZ$ and equals the monopole's magnetic charge.

\item 
At low energy  it  is possible  to separate  the solitonic 
``particle'' degrees
of freedom from other degrees of freedom (such as  radiation)
 in the field theory. One may then  truncate the 
theory and model  soliton dynamics by the time evolution of finitely many
collective coordinates. For $SU(2)$ monopoles in the Prasad-Sommerfield
limit  the moduli spaces of static soliton solutions may be used
as collective coordinates. The moduli space of magnetic charge $K$ 
monopoles is denoted $M_K$. An elementary but crucial property is 
that  the dimension of $M_K$ increases linearly with $K$
(in this case dim$M_K =4K$). Thus the $K$-particle moduli space has enough
degrees of freedom to allow for independent motion of the $K$ particles.

\item
The unbroken gauge group  acts smoothly on  the moduli spaces.
Semi-classically,  quantum  states are realised as wavefunctions 
on the moduli space and it  follows that they can be organised
into irreducible representations of the unbroken gauge group. 
In the case at hand where  the unbroken gauge group is $U(1)$
the representations are labelled by a single integer $N$ which physically
corresponds to the electric charge. General dyonic quantum states 
are therefore labelled by two integers: $K$ characterising  the magnetic
and $N$ characterising the electric charge.

\item
Although magnetic and electric charges  have a very different mathematical
status (the former being a topological charge, the latter being a Noether
charge) one can envisage an action on the dyonic states 
which exchanges  the integers $K$ and $N$. This is what happens in the
electric-magnetic duality conjecture of Montonen and Olive. 

\end{enumerate}

If one goes through the above list and checks which of these
properties still holds   for monopoles in theories with 
non-abelian unbroken gauge group one encounters some surprises. 
We consider the simplest case,
where the gauge group $SU(3)$ is broken to $U(2)$, and formulate
some pertinent questions.

\begin{enumerate}

\item 
Here the situation is analogous.
Monopoles are   topologically classified  in terms of 
$\Pi_2(SU(3)/U(2))=\bZ$. Thus there is again  an integer $K$ which 
specifies the monopole's topological charge and which one can interpret
as a particle number.

\item
The moduli spaces are still 
labelled by the topological magnetic charge, but in addition
they are stratified. For topological charge $K\geq 0$ there are $[K/2]+1$
strata (square brackets denote the integer part), with each stratum
corresponding to one of $[K/2]+1$ distinct ways in which $K$ monopoles can
be put together. Different strata have different dimensions:
the largest has dimension $6K$, but all others have lower dimension.
What is the physical significance of the strata?

\item
The action of the unbroken gauge group $U(2)$ on the moduli space
 depends on the stratum. This is a manifestation of the effect 
first noted by Abouelsaood in \cite{Abou,Abouu}  
that dyonic excitations  of monopoles
in theories with non-abelian magnetic gauge group do not generally
fall into representations of the unbroken gauge group. 
Instead there is a subtle dependence of the ``electric group''
on the magnetic charge in such theories. 
What is the algebraic structure  behind this interplay?

\item
With topological magnetic charge given by a single integer, and 
the electric charge sometimes given by $U(2)$ representations and
sometimes (as we shall see) by  a $U(1)\times U(1)$ representation
it is clear that electric-magnetic duality cannot be realised simply as 
the exchange of magnetic and electric properties. How should one 
formulate it instead?

\end{enumerate}

The goal of the rest of the talk is to provide a conceptual framework
for addressing and partly answering these questions.

\section{$SU(3)$ monopoles and their moduli spaces}
A monopole solution of 
 $SU(3)$  Yang-Mills-Higgs theory with coupling constant $e$
 in the Bogomol'nyi
limit is a   pair $(A_i,\Phi)$
of a SU(3) connection $A_i$ and an adjoint  Higgs field $\Phi$
on ${\bR}^3$
satisfying 
the Bogomol'nyi equations
\bea
\label{bog}
D_i\Phi = B_i,
\eea
where $D_i=\partial_i +  e A_i$ is the  covariant derivative and $B_i$
is the non-abelian magnetic field constructed from $A_i$.
Appropriate boundary conditions have to be imposed to ensure that the
energy is finite, and one also demands that the Higgs field has
the following form along the positive $z$-axis:
\bea
\Phi(0,0,z) = \Phi_0 - {G_0 \over 4 \pi  z} +{\cal  O}({1\over z^2}),
\eea
where   $\Phi_0$ is a constant non-vanishing element of the Lie algebra of 
$SU(3)$, chosen to lie
in the Cartan subalgebra of diagonal traceless matrices.
$\Phi_0$ determines the symmetry breaking pattern. 
If it has three distinct eigenvalues  the symmetry is broken
maximally to $U(1)\times U(1)$. If two eigenvalues coincide the
symmetry is broken minimally to $U(2)$; this is the case of interest
here.

The Bogomol'nyi equation relates the Higgs field to the magnetic
field and shows that one may interpret  $G_0$ as the vector 
magnetic charge of the monopole. 
 It  also follows from the Bogomol'nyi equation (\ref{bog})
 that  $G_0$ commutes with $\Phi_0$ \cite{Weinberg} so that one may
rotate $G_0$ into the Cartan subalgebra. According to the generalised
Dirac condition  \cite{EW},\cite{GNO}, 
$G_0$  has to lie on the dual root lattice after that rotation,
which means that  it has the form
\bea
\label{dirac}
G_0= {2 \pi\over e} \mbox{diag}(m_1,m_2-m_1,m_2)
\eea
for integers $m_1$ and $m_2$. This condition is usually  derived without
reference to the symmetry breaking pattern.
In the case of maximal symmetry breaking one can show that both
the integers  appearing in (\ref{dirac})
have a topological significance.
In the case of minimal symmetry breaking, however, there is only
one topological charge. What is the relevance of the second integer?

In the Bogomol'nyi
limit the answer to this question was given in the recent mathematical
literature, particularly in the work of Donaldson and Murray. Extending
Donaldson's work relating magnetic monopoles to  rational
maps, Murray showed in \cite{Murray} 
 that the integer which lacks 
 a topological interpretation characterises holomorphic
properties of the  monopole. Thus one can say that in general the
vector  magnetic magnetic charge has topological and holomorphic components.
One important difference between  the two sorts of  charges  is revealed 
by  the action of the unbroken gauge group  on $G_0$. 
In general this generates an
orbit, but  more precisely it is the holomorphic components 
which sweep out a non-trivial  orbit while the topological
components remain invariant. In  \cite{BS}
 we christened  these non-trivial orbits
`magnetic orbits' and pointed out  that  one can usefully
translate the holomorphic charges defined by Murray into  numbers
characterising these  orbits.

In the theory at hand  the magnetic orbits are 
the cosets $U(2)/(U(1)\times U(1))$ and thus have the topology of a two-sphere.
As explained in \cite{BS} these two-spheres all have their centres on the
one-dimensional lattice $\{K\Phi_0\}_{K\in {\bf Z}}$
 in  the Lie algebra of $SU(3)$.
While  the topological charge specifies 
the position of the centre of the two-sphere  the holomorphic charge
specifies  its radius.  As a consequence 
of the Dirac condition  the allowed values for the radius 
 are (positive) half-integers.
Thus one may picture the magnetic properties of monopoles in YMH theory
with gauge group $SU(3)$ broken to $U(2)$ as two-spheres 
in the Lie algebra of $SU(3)$  with quantised
radii and and centers. Note that these orbits intersect the 
Cartan subalgebra on the lattice  defined by the 
Dirac condition (\ref{dirac}).
 Murray showed that solutions of the 
Bogomol'nyi equations only exist for certain holomorphic charges;
translated into our language his results say 
  that  solutions exist for  magnetic orbits with 
arbitrary centers $K$ but only for  radii  $k$
 in the set $\{0,1,2, ..., |K|/2\}$ if $K$ is even, and in the set
$\{1/2, 3/2, ..., |K|/2\}$ if $K$ is odd.

Monopole moduli spaces are defined as the set of all monopole solutions
of a given  topological charge $K$, divided by the group
\bea
{\cal G}_0 = \bigl\{
g:  {\bR}^3 \rightarrow SU(3)\, \big \vert \lim_{z \rightarrow \infty}
 g(0,0,z) = \hbox{id}\bigr\}
\eea
 of  framed gauge
transformations.  Murray showed  that the moduli spaces
are further subdivided into strata, labelled by the 
holomorphic charges and  each  containing
all monopoles of  the given holomorphic charge.
In our picture the magnetic orbit's centre
labels the moduli spaces and the magnetic  orbit's radius labels  the 
strata. 
 Thus  we denote the strata  by
$M_{K,k}$. The different strata for given $K$ 
generally have different dimensions, but they form
part of one connected space.

So far we have only described the moduli spaces as sets,
but it is only when we  induce more structure from
the field theory onto these spaces 
 that we can use them to answer physical questions.
The first question we want to address is the  question of 
 how the exact symmetry group is realised
in the various magnetic sectors. Seminal papers by Abouelsaood
 \cite{Abou,Abouu} and  
Nelson and Manohar \cite{NM}  made it clear that 
``electric'' excitations of a monopole with given vector magnetic 
charge do not, as one might naively expect, fall into representations
of the exact group $U(2)$ but  only form representations of the
subgroup  of $U(2)$ which commutes with  the vector magnetic charge
(the centraliser subgroup). The basic reason for this is that 
infinitesimal deformations which  change the magnetic field at
infinity do not satisfy Gauss' law and are therefore 
physically ruled out.
In the $SU(3)$ example these results imply  that  electric 
excitations of monopoles  with ``purely topological''
vector  magnetic charges  ($k=0$) form $U(2)$ representations
while for all  other  vector  magnetic charges the electric excitations
only carry representations of a $U(1)\times U(1)$ subgroup of 
$U(2)$. Moreover, in
the latter case one of the $U(1)$  groups depends on 
 the vector magnetic charge.
In other words,  monopoles in this sector may be charged with respect
to different $U(1)$ subgroups.

How are these  physical facts reflected in the moduli spaces?
For $K=1$  there is only one allowed
value of the orbit radius, so the   moduli space only
has  one stratum, namely $M_{1,1/2}$.
 Murray showed that this space   is six-dimensional
and that it   has the topology ${\bR}^3 \times S^3$. The first factor
parametrises the monopole's position and in \cite{BS}
 it is further explained that
the $S^3$ part should physically be interpreted by thinking of 
it as Hopf-fibred over a two-sphere. The base space of this fibration 
is  the magnetic two-sphere described earlier
 and the  fibre is an electric circle, 
 familiar from $SU(2)$ monopoles.
Motion on this circle is physical and  gives the monopole electric charge, but
motion tangent to the base space is forbidden because it violates
Gauss' law. The fibration
thus captures and makes precise
 the interplay between magnetic and electric properties
suggested by the work of Abouelsaood {\it et al}. A point on the 
magnetic two-sphere specifies the monopole's magnetic charge,
and only the $U(1)$-rotations about this magnetic direction
are permissible electric excitations. 

Moving on to topological 
charge $K=2$  we find a  moduli space with two strata. The space
$M_{2,0}$, called the
large stratum, contains monopoles with purely topological
charge. It  is a  12-dimensional
smooth manifold and was studied by Dancer in a series of
papers \cite{Dan,Danc,Dance}.
The whole unbroken gauge group $U(2)$ acts smoothly on this space.  
The stratum $M_{2,1}$, called the small stratum,  
corresponds to a  magnetic orbit
of radius  one and is 10-dimensional; like the 
space $M_{1,1/2}$ it is fibred over the magnetic orbit and 
in a  fibre over a given magnetic charge
 only that charge's  centraliser subgroup of $U(2)$ 
acts smoothly.

Apart from the  group action of  the unbroken gauge group, 
a further physically important property of the moduli spaces is the metric
they inherit from the kinetic energy of the field theory. 
No metric can be defined on those parts of the moduli space 
whose tangent vectors violate Gauss' law, namely the magnetic
orbits. In general, the metric structure and the group action
of $U(2)$ can be summarised as follows.  Those strata of 
the moduli spaces which are labelled by magnetic orbits of
zero radius are smooth manifold with hyperk\"ahler metrics
and smooth $U(2)$ actions. All other strata have the structure
of a fibre bundle over the magnetic orbit; the fibres of this
fibration are smooth hyperk\"ahler manifolds  which only
permit smooth actions of that $U(1)\times U(1)$ 
 subgroup of $U(2)$ which
leaves the magnetic charge labelling the fibre invariant.

\section{Dyonic quantum states and the emergence of $U(2)\ltimes{\bR}^4$}

In the case where the unbroken gauge symmetry is abelian  the 
following semi-classical bosonic  quantisation scheme 
for monopoles  in the BPS limit has been standard in the literature. 
The Hilbert space of states is taken to be 
 the space  of (square-integrable) wavefunctions
on the monopole  moduli space and the covariant 
Laplacian on the moduli space plays the role of the quantum  Hamiltonian.
This prescription can be extended to  a supersymmetric situation.
If one thinks of the monopoles as classical bosonic solutions in 
 $N=4$ supersymmetric Yang-Mills
 theory, the quantum mechanical model for their motion 
is $N=4$ supersymmetric quantum mechanics on 
the moduli space: the Hilbert space  is the space of all
(square-integrable)  forms on 
the moduli space and the Hamiltonian is the Laplacian acting on forms.
An important consistency requirement for this quantum mechanical
model is the hyperk\"ahler property of the metric.

In theories with unbroken non-abelian symmetry the above
scheme has to be modified. The fibration of the strata is crucial.
In each stratum   the above quantisation 
scheme can be applied to the wavefunctions on the fibres (as remarked, these
have hyperk\"ahler metrics). By contrast points on the magnetic orbits
  (the base spaces of the fibration) serve as labels of superselection  sectors
of the theory. Since a point on the magnetic orbit also specifies which
subgroup  of the unbroken gauge group  can be implemented physically as 
the electric group one arrives at the conclusion that dyonic 
quantum  states are labelled by a point on the magnetic orbit together with a
representation of the centraliser subgroup of that point. In the
YMH theory with gauge group  $SU(3)$ broken to $U(2)$ we therefore
have  the following labelling of dyonic states. Writing $K$ for the topological
magnetic charge as before and $k$ for the radius of the magnetic sphere,
we specify the magnetic  charges by giving $K$ and a vector $\bf k$
on the magnetic sphere (and thus of length $k$). If $k=0$ the electric
group is the full group $U(2)=(SU(2)\times U(1))/{\bf Z}_2$.
States in $U(2)$ representations are labelled by three integers $j,m,N$,
with $j=0,1/2, ..$ and $m\in\{-j,-j+1, ..., j-1,j\}$ 
specifying a state in a  $SU(2)$ representation  and $N$ specifying
 a $U(1)$ representation;
the ${\bf Z}_2$ identification requires  that $N+ 2j$ be even.
Thus dyonic states on the strata with trivial magnetic orbits are of 
the form
\bea
\label{state}
|K,0;N,j,m\rangle.
\eea
Introducing an explicit parametrisation of $U(2)$ in terms of 
a $U(1)$-angle $\chi \in [0,2\pi)$ and Euler angles $(\alpha,\beta,\gamma)$
for $SU(2)$ one can represent the  above state as a function on $U(2)$,
using the Wigner functions $D_{ms}^j$ on $SU(2)$:
\bea
\langle \chi , \alpha, \beta, \gamma |K,0;N, j,m \rangle=
e^{iN\chi} D_{ms}^{j*}(\alpha,\beta,\gamma)
\eea
(Different values  of $s$ lead to equally valid realisations
of the state (\ref{state})). The angles 
$(\chi , \alpha, \beta, \gamma)$ explicitly appear in the parametrisation
of large strata such as $M_{2,0}$, so
the above formula shows how to realise  dyonic states in that
sector as wavefunctions on the  moduli space.

In the strata with magnetic orbits of radius $k>0$, the magnetic charge
is specified by giving $K$ and ${\bf k}$ as defined above, while 
 the electric group is $U(1)\times U(1)$ (with the second factor
being the centraliser group of ${\bf k} $ in $ SU(2)$)
 whose representations are labelled by one integer
$N$ and one half-integer $s$. Thus dyonic quantum states in these
strata  can be written as 
\bea
\label{dstate}
|K,{\bf k};N,s\rangle.
\eea
In particular for a single monopole, $K=1$ and $k=1/2$, we have the 
additional constraint $N=2s$, and can represent the above state
as a function on $M_{1,1/2}$. 
Using again Euler angles $(\alpha,\beta,\gamma)$ for the $S^3$
part of that space, and parametrising  the direction of ${\bf k}$
by spherical coordinates $(\hat \beta,\hat \alpha)$ (so that 
${\bf k} = (\sin\hat\beta \cos\hat\alpha,\sin\hat\beta \sin\hat\alpha,
\cos\hat\beta)$) one finds
\bea
\langle\alpha,\beta,\gamma|1,{\bf k},2s,s\rangle = \delta(\cos \beta
-\cos \hat \beta)\delta(\alpha -\hat \alpha)e^{i s \gamma}.
\eea

A key observation of \cite{BS} is that all these  dyonic 
states can profitably
be interpreted as states in representations of the semi-direct product
$U(2)\ltimes {\bR}^4$.  The non-trivial part of this group is the double
cover of the three dimensional Euclidean group,  in whose representation
theory the interplay between orbits and centraliser representations is
familiar, albeit in a different guise. In that context irreducible 
representations $V_{k,s}$ are labelled by the magnitude $k>0$ of the momentum
vector $\bf k$ and a half integer $s$
specifying the helicity, which is a representation of the $U(1)$ subgroup which
leaves a specified momentum vector  invariant.
These representations are infinite dimensional, and  in it 
 translation and helicity
eigenstates are labelled by a momentum vector ${\bf k}$  of length $k$ and 
the half-integer $s$. If ${\bf k}=0$ the representation spaces are 
written  $V_{0,j}$ and isomorphic to the usual $(2j+1)$-dimensional 
spin $j$ representations of $SU(2)$. Here I  have chosen a notation
that makes the correspondence between the different physical situations
evident; for explicit formulae which also include the $U(1)$ part 
of $U(2)$ I refer the listener to \cite{BS}.

Interpreting dyonic states as elements of 
 $U(2)\ltimes {\bR}^4$ representations 
answers the questions posed under points 2. and 3. in our initial list.
The labels of the different strata of the moduli spaces now gain
a group-theoretic interpretation, and quantum states on 
different strata can be combined according to the Clebsch-Gordan
coefficients of $U(2)\ltimes {\bR}^4$. In particular one can now 
understand how a  dyonic state in the large stratum $M_{2,0}$
 (which carries a $U(2)$ representation) can be a tensor product
 of two dyonic states of two single monopoles
(which  only carry  $U(1)$ electric charges). The trick that made this
possible was to interpret both magnetic and electric properties 
as representation labels of one algebraic object. I should also 
emphasise  that  this trick depends crucially on the  inclusion of 
the full magnetic orbit in the discussion. Restricting attention
to a particular magnetic charge, or (as many authors have done)
to the Weyl orbit of a particular charge, makes a consistent fusion
algebra of non-abelian dyons impossible.

\section{Discussion and outlook}

While much  progress in solving our initial puzzles could
be made by interpreting non-abelian
 dyons in the theory discussed here  as carriers of  $U(2)\ltimes {\bR}^4$
representations, a number of open questions remain. The first
concerns the Dirac quantisation of the magnetic orbits. While this 
is a fundamental fact from the point of view of monopole
physics it has to be  imposed artificially  in the representation
theory of $U(2)\ltimes {\bR}^4$. Also, to respect this condition
when multiplying dyonic states in a tensor product one has to 
require that the magnetic charge vectors ${\bf k}_1$ and ${\bf k}_2$
of the two dyons to be multiplied are either parallel or anti-parallel.
While these conditions can be imposed consistently  they suggest that 
there is a further algebraic object,  related to semi-direct products
but in some way more restrictive, whose representation theory incorporates
the Dirac conditions from the start. 

Similarly (and perhaps relatedly) more work is required to 
answer the question posed  under point 4. of  our list, namely the proper
formulation of duality in Yang-Mills theory with non-abelian
unbroken gauge group. The organisation of dyonic states into
representations of semi-direct product groups allows one to formulate
this question  in sharper language. In \cite{BS} a natural candidate
for a electric-magnetic duality
transformation is discussed which acts on the representations $V_{k,s}$
by exchanging the magnetic orbit label $k$ with the electric centraliser
label $s$; since $k$ is quantised by the Dirac condition such an exchange
makes sense. However, while this appears to be satisfactory as long as both
$k$ and $s$ are non-vanishing, it is problematic when $s=0$.
For $k=1/2$ for example  this prescription would 
then relate the {\it infinite}-dimensional representation
$V_{1/2,0}$ containing all  single monopole states  with the purely
electric {\it  two}-dimensional representation $V_{0,1/2}$. Since duality 
is supposed to relate degrees of freedom which are in a suitable sense
equivalent, such a mismatch of dimensions is not acceptable.
It seems that  further conceptual progress  is required
before we  fully understand non-abelian dyons.

\vspace{2cm}

\noindent {\bf Acknowledgments} 

\noindent The work reported here was done in collaboration with Sander Bais
and financially supported by a Pionier fund of the Nederlandse Organisatie
voor Wetenschappelijk Onderzoek. I thank the organisers for
organising a stimulating workshop and for  inviting me 
to speak.

\end{document}